\documentclass[aps,prl,twocolumn,groupedaddress]{revtex4}

\usepackage{graphicx}
\usepackage{amsmath}
\usepackage{color}
\usepackage{epstopdf}

\begin{document}
\title{Terahertz cascades from nanoparticles}
\author{K.B. Arnardottir}
\email{kristinbjorga@gmail.com}
\author{T.C.H. Liew}
\email{tchliew@gmail.com}
\affiliation{Division of Physics and Applied Physics, School of Physical and Mathematical Sciences, Nanyang Technological University, 21 Nanyang Link, 637371 Singapore}

\date{\today}

\begin{abstract}
In this article we propose a system capable of THz radiation with quantum yield above unity. The system consists of nanoparticles where the material composition varies along the radial direction of each nanoparticle in such a way that a ladder of equidistant energy levels emerges. By then exciting the highest level of this ladder we produce multiple photons of the same frequency in the THz range. We demonstrate how we can calculate a continuous material composition profile that achieves a high quantum yield and then show that a more experimentally friendly design of a multishell nanoparticle can still result in a high quantum yield.
\end{abstract}

\maketitle

The lack of terahertz (THz) frequency electromagnetic transitions in typical materials allows THz frequency radiation to pass through many materials freely without causing ionization, which is well-known to lead to a variety of applications ranging from medical imaging and security screening to supply chain management. At the same time, the limited interaction between THz radiation and many materials limits the possibilities for generating THz radiation for those same applications that require transparency.

As semiconductors have grown to become the system of choice for the generation of optical radiation, they have also been seen as prime candidates for generating THz radiation, albeit with quite different physical mechanisms. Typically one needs to find and excite a low energy THz transition available in the system, and a good place to look is at the lowest energy semiconductor excitations, namely excitons. In the same way that a hydrogen atom can emit at optical wavelengths, the larger excitons can emit THz radiation by making a 2p-1s transition~\cite{Kavokin2012,Kaliteevski2014,Menard2014}. Other methods using excitons have been based on coupling to light resulting in the THz frequency splitting of hybrid light-matter modes (exciton-polaritons)~\cite{delValle2011,Savenko2011}, which can exhibit transitions in asymmetric systems~\cite{DeLiberato2013,Huppert2014,Barachati2015} or via the mixing of different varieties of excitons~\cite{Kyriienko2012,Kristinsson2013}.
There are also proposals which do not use excitons to achieve THz radiation, such as the recently realized system of a low-cost metallic heterostructure \cite{Wu2017} as well as the recently proposed system of an ensemble of asymmetric quantum dots \cite{Vanik2017}.

The aforementioned methods are highly promising as they appear in compact structures (quantum wells). They also fall into a broad class of methods that can be summarized as aiming to convert a relatively high energy quantum (such as an optical photon) into a much lower energy THz photon. In the ideal case this is achieved with unit quantum yield. In terms of energy yield the generation of THz radiation then becomes expensive given the large imbalance of optical energies put into the system and THz energies coming out. To do better than unit quantum yield one needs to make use of cascaded processes~\cite{Kohler2002,Tzimis2015,Belkin2015,Liew2013}, where a single input quantum can generate many THz photons. Indeed devices based on quantum cascade lasers achieve the very best efficiencies in the THz field~\cite{Liu2013}, while they are typically considered bulky systems.

In this work we aim to merge the concept of THz cascades based on fermionic transitions in semiconductor heterostructures with the concept of using artificial atoms. Instead of using excitons as artificial atoms we consider semiconductor nanoparticles, which lie among the most compact of man-made systems. It has been suggested that the alloy composition of such nanoparticles can be varied spatially \cite{Baimuratov2014,Aktas2008}. We show that in principle this leads to the possibility of engineering multiple transitions with the same THz range frequency in a single nanoparticle. To obtain a suitable alloy composition profile, we introduce a numerical optimization algorithm, based on a form of gradient descent. We calculate the optimized THz transition rates for a typical GaAlAs based system. Considering a collection of nanoparticles in a compact cavity, quantum yield greater than unity is readily obtained.

Even though these kinds of particles have not yet been experimentally achieved, we assume that the method of fabricating core/shell semiconductor quantum dots\cite{Chang2007,Fang2009} can be expanded upon to add  more layers to make larger particles. Currently, dots with up to four layers have been demonstrated \cite{Zhang2009,Yang2014}.

\begin{figure}
\includegraphics[width=0.5\textwidth]{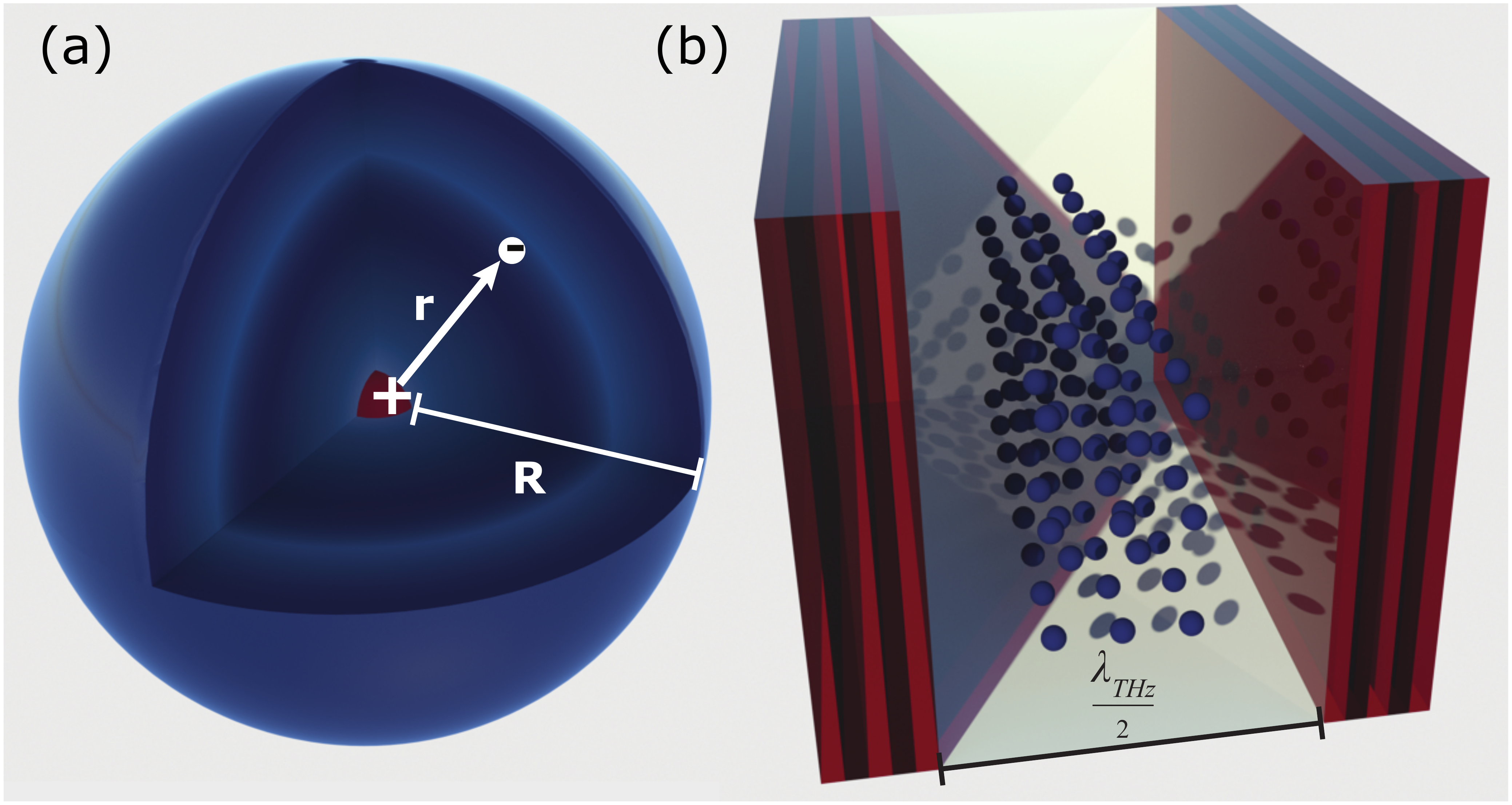}
\caption{a) We consider a nanoparticle with a hydrogenic impurity in the center and whose material composition changes along the radius $r$. This means that an ion of charge $+e$ is located in the center and electrons in the dot will feel a modified Coulomb potential. Our goal is to find parameters that can result in a ladder of equidistant energy transitions that result in radiation in the THz range.
b) When multiple such particles are embedded into a cavity resonant to the resonant frequency of the ladder of transitions, i.e. $\lambda_{THz}=2\pi c/\omega_{THz}$ where $\hbar\omega_{THz}$ is the energy of the transitions, we can enhance the transition between the chosen levels through stimulated emission.} \label{fig:sketch}
\end{figure}

{\it Model.---} The system we consider consists of a spherical nanoparticle of radius $R$ with a hydrogenic impurity in the center \cite{Chuu1992,Baimuratov2014,Hsieh2000}, as seen in Fig. \ref{fig:sketch}. The material of the nanoparticle varies along the radial direction $r$. 

One way to achieve this physically would be to consider a Ga$_{1-x}$Al$_x$As nanoparticle where the aluminium concentration $x$ varies along the radial direction. For simplicity, in both calculations and fabrication, we take the particle to be spherically symmetric such that $x(r)$ has only radial dependence.
The inhomogeneity means that the effective mass of electrons in the dot, $\mu$, and the electric permittivity, $\epsilon$, will both vary along $r$. In a Ga$_{1-x}$Al$_x$As this leads to a linear dependence: $\epsilon(x)=12.9(1-0.22x)\epsilon_0$ and $\mu(x)=0.063(1+1.32x) m_0$, where $m_0$ is the free electron mass \cite{Aktas2008}.
This means that the Hamiltonian will depend on the function $x(r)$:
\begin{equation}
\label{eq:H}
H[x(r)]=-\frac{\hbar^2}{2}\nabla\frac{1}{\mu(x(r))}\nabla
+\frac{e^2}{4\pi}\int_r^R\frac{dr'}{\epsilon(x(r'))r'^2},
\end{equation}
where the wavefunction, $\psi$, must fulfil the boundary condition $\psi(r)=0$ for $r\geq R$.

Now, we want to find a function $x(r)$ which leads to a ladder of transitions of a particular frequency in the THz range. To do this we need to fix the nanoparticle radius $R$, the coveted frequency $\omega_{THz}$ and a set of levels to be shifted, $\{L_i\}$. Note that these levels need to obey the selection rule $\Delta \ell=\pm1$, where $\ell$ is the orbital angular momentum, so as to give a possibility to excite the highest state easily.

To find $x(r)$ we make use of variational calculus. The energy of level $i$ can be written as a functional of $x(r)$:
\begin{equation}
E_i=\int \underbrace{\psi_i^* H[x(r)]\psi_i}_{L(r,x(r))} d^3r,
\end{equation}
where $\psi_i$ is the wavefunction of level $i$.
If $x$ is varied by $\delta x$, that is $x\rightarrow x+\delta x$, the energy level will change by
\begin{equation}
\delta E_i=\int\frac{\delta L}{\delta x}\delta x d^3 r
\end{equation}
We can use this equation to choose the varying function $\delta x$ that leads to the shift of energy levels we want. We take
\begin{equation}
\delta x=\beta C \Delta E \frac{\delta L}{\delta x},
\end{equation}
where $C^{-1}=\int\left(\frac{\delta L}{\delta x}\right)^2d^3r$ is a normalization constant and $\Delta E=E_{ideal}-E_i$ where $E_{ideal}$ is the energy we want to shift level $i$ to coincide with. The factor $\beta\leq 1$ can be tuned to keep the variation small for each step of the iteration so that the system gradually relaxes to an optimum condition. 

This transformation is then repeated for each level in the chosen level set $\{L_i\}$. This algorithm can then be iterated until each energy level $E_i$ is sufficiently close to the ideal energy. The calculations are done numerically in the basis of the radial coordinate $r$ where the integral in the last term of Eq. (\ref{eq:H}) becomes a multiplication by a triangular matrix and differentiation is calculated using first-order divided difference, where the symmetric difference quotient was chosen for first order differentiation.

After we have found parameters that give agreeable results we can then use the final wavefunctions to calculate the dipole moment corresponding to available transitions in the system. We then set up rate equations coupling the occupations of the levels as well as the THz photon mode. By introducing a finite lifetime for the photon mode we can simulate putting the nanoparticle inside a cavity resonant to the THz frequency of the photons. Furthermore, we can modify the rate equations to have multiple particles inside the cavity.

{\it Results and discussion.---} We choose the radius $R$ so that the transition energies between the level set $\{L_i\}$ are as close to the coveted energy $\hbar\omega_{THz}$ as needed.
When choosing the sets of levels to consider we limit ourselves to the case where the highest level is a $p$ state (angular momentum $\ell=1$) so that it can be excited by a laser from the ground state.
Once the values of $R, \omega_{THz}$ and the level set have been chosen, we apply the aforementioned iterative approach to find a function $x(r)$ that modifies the energy levels to approach an equidistant ladder.

\begin{figure}
\includegraphics[width=0.5\textwidth]{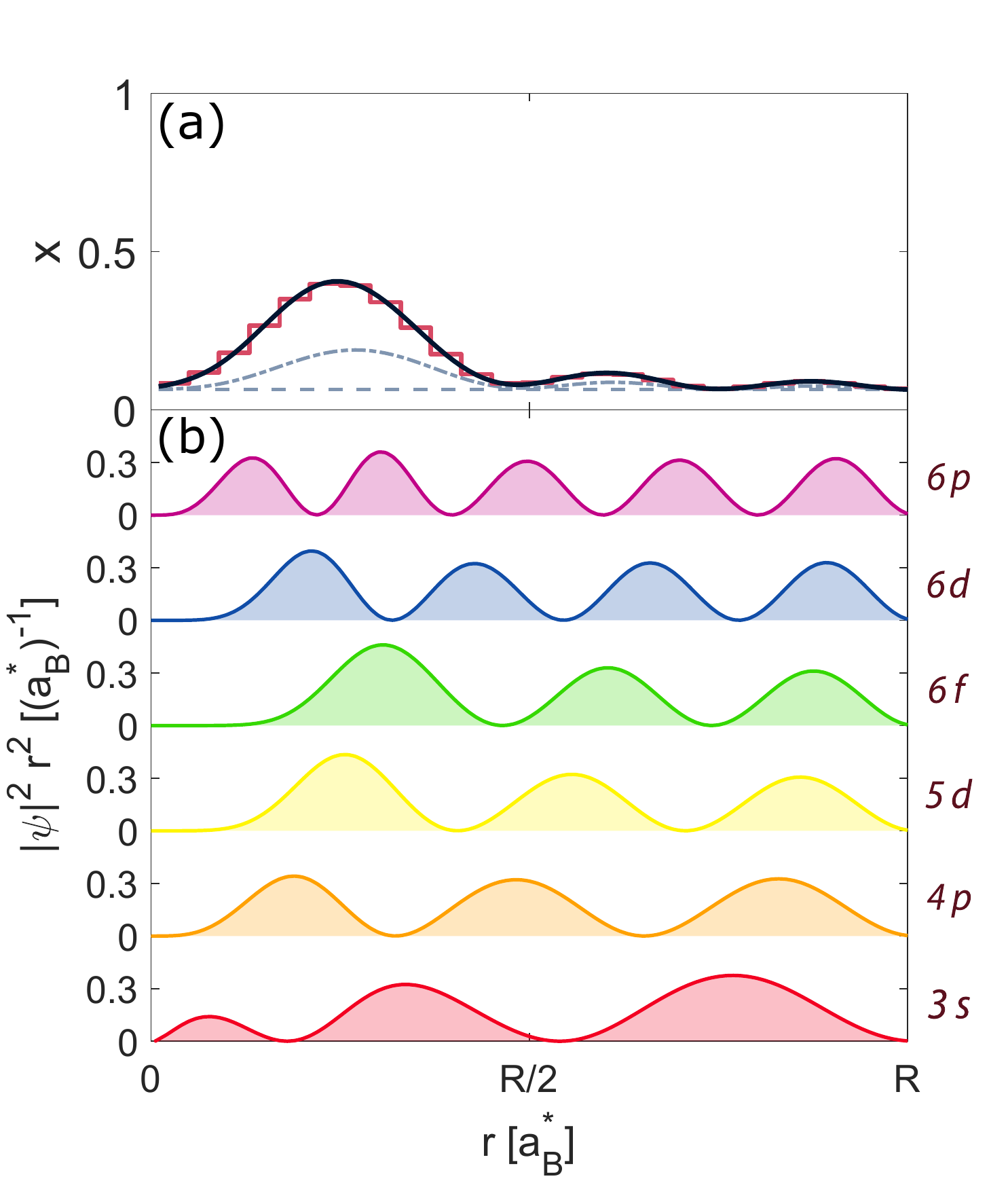}
\caption{(Color online) (a) The upper graph shows the function $x(r)$, which describes the Al concentration in a GaAlAs nanoparticle. The lighter, dashed line corresponds to the initial guess for $x$. The final result, shown by the dark, solid line, was obtained after only 10 iterations, at which point the convergence was deemed high enough.
This $x(r)$ gives rise to an equidistant ladder of transitions for previously chosen energy levels. The step function plotted in solid light red is an approximation of the former curve which should be easier to fabricate. The number of steps is 25.
(b) The lower part of the graph depicts the radial part of the wave functions of the levels. The parameters used were for a Ga$_{1-x}$Al$_x$As nanoparticle of radius $R=$63 nm, where the energy difference between the selected levels corresponds to radiation frequency of $\omega_{THz}=1.1$ THz. We note that Ga$_{1-x}$Al$_x$As has a direct bandgap for $x<0.44$ \cite{Allali1993}.}\label{fig:x}
\end{figure}

The function $x(r)$ calculated from $R=63$ nm, $\omega_{THz}=$1.1 THz and $\{L_i\}=\{3s, 4p, 5d, 6f, 6d, 6p \}$ after 10 iterations can be seen in Fig. \ref{fig:x}a, along with the initial condition (plotted as a dashed line) and the result after 5 iterations (plotted as a dashed-dotted line). Note that the level set considered for optimization does not include the ground state $1s$. This is intentional, since we want to make sure that the system can be pumped by an infrared source; this requires a large gap between the 1s and other states. Our simple iterative procedure does not guarantee finding a global optimum to the problem at hand, but it can efficiently find a local optimum for a well-chosen initial condition. The graph also shows a piecewise approximation of the function which should be easier to fabricate experimentally, plotted in solid light red. This approximation is a step function with 25 steps, and will lead to a less exact ladder of transitions. This means that we need to take a more lossy cavity, where the cavity photon mode has a larger linewidth.

\begin{figure}
\includegraphics[width=0.45\textwidth]{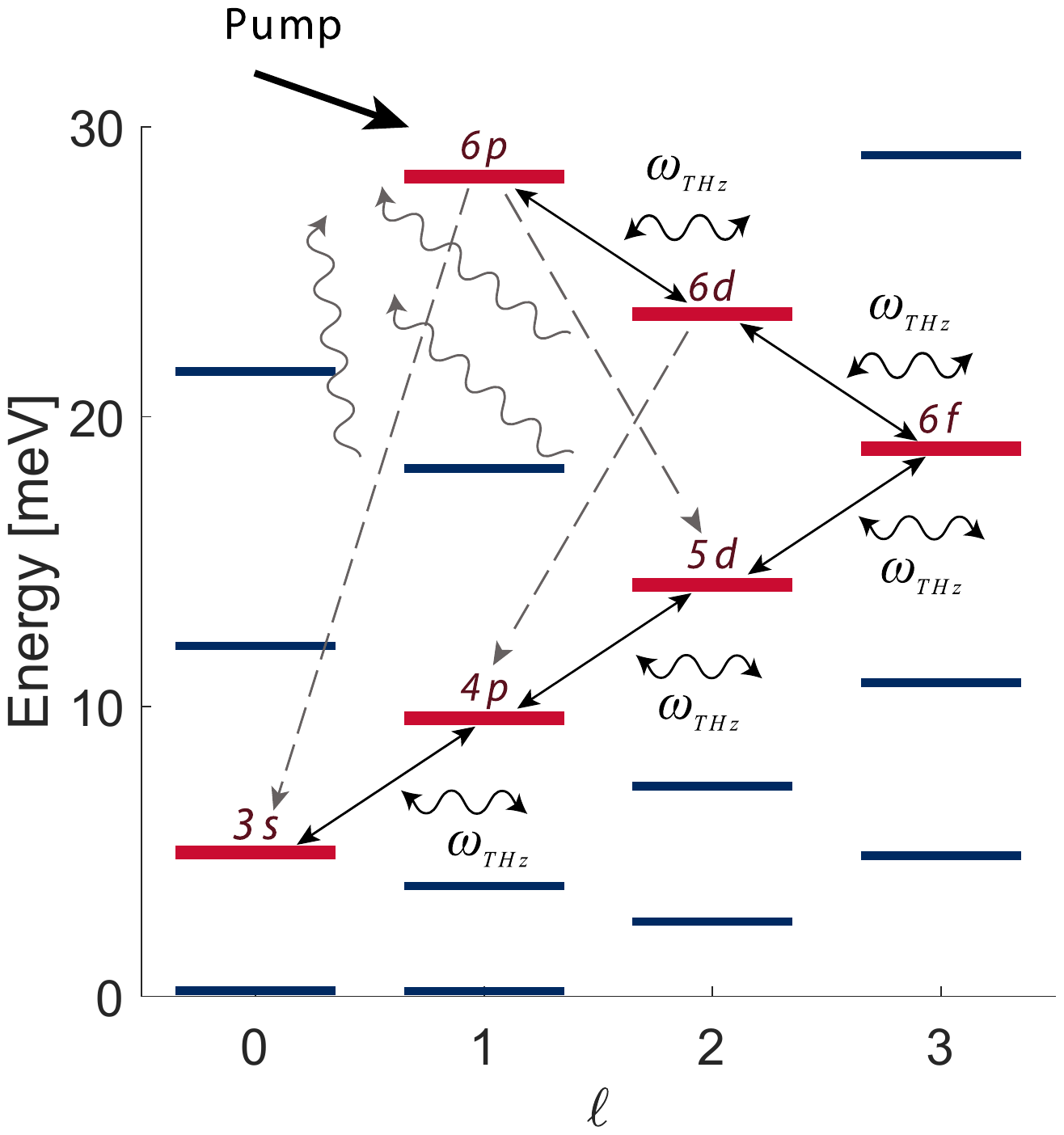}
\caption{(Color online)The energy levels in the nanoparticle after using the recursive algorithm discussed in text. The chosen set of levels that make the equidistant ladder are denoted by thick red lines while other levels in the system are thinner, blue lines. The transitions shown by black, solid arrows are all of a single transition energy corresponding to$\omega_{THz}=$1.1 THz (or 4.6 meV), while the gray, dashed arrows correspond to transitions of different energies and lead to a loss in the system. Other transitions are forbidden due to selection rules.}
\label{fig:levels}
\end{figure}

Fig. \ref{fig:x}b shows the radial parts of the wavefunctions corresponding to the energy levels forming the ladder. For clarity the intensity of the wavefunctions scaled by $r^2$ is plotted ($|\psi|^2r^2$ is proportional to the intensity of a given state at particular $r$ integrated over the angular coordinates). 
Fig. \ref{fig:levels} shows the modified levels for the same parameters. The thick, red lines are the chosen set of levels which are to form the equidistant ladder. The black, solid arrows correspond to transitions resonant with the coveted frequency $\omega_{THz}$ while the gray, dashed arrows are possible transitions which result in radiation of a different frequency and therefore lead to loss in our system. We assume an external pump that excites electrons in the ground state to the top of our ladder, the $6p$ state (the 1s-6p energy gap corresponds to 60 THz for our parameters, which lies in the infrared region). After the electron has been relaxed to the $3s$ state it will decay to the ground state through lower energy states such as $1p, 2p$ and $3p$. The transition rates between these states and the $3s$ state on one hand and the ground state on the other hand will determine the lifetime of the $3s$ state.

We also need to take note of the working temperature, since the energy transitions are generally smaller than the thermal energy of room temperature. We have to make sure that in the absence of pumping the ground state is dominantly occupied, which breaks down if the thermal energy is larger than the transition between the ground state and the next excited state. For our parameters this difference is 7.7 meV, which corresponds to a temperature of 90 K. We therefore need to make sure that the working temperature is lower than 90 K, which can be achieved using liquid nitrogen.

Using the eigenfunctions of the Hamiltonian in Eq. (\ref{eq:H}) we can calculate the transition rates between the possible transitions. The transition rate between two states labelled with $i$ and $j$ respectively, can be written \cite{CohenTannoudji1997}
\begin{equation}
\label{eq:W}
W_{ij}=\frac{\alpha^4c}{6 a_B^*}\left(\frac{\Delta E_{ij}}{Ry^*}\right)^3\left(\frac{|\langle i|\mathbf{r}|j\rangle|}{a_B^*}\right)^2,
\end{equation}
where $\alpha$ is the fine structure constant, $a_B^*$ and $Ry^*$ are the effective Bohr radius and Rydberg energy for GaAs and $\Delta E_{ij}$ is the energy difference between the levels $i$ and $j$.

With the transition rates known we can write out rate equations for the modes in our system. With $n_i$ denoting the occupation of level $i$ and $n_{\gamma}$ the occupation of the photon mode, we have

\begin{align}
\label{eq:rate}
\frac{dn_i}{dt}&=\sum_{j>i}W_{ij}\left[n_j(1-n_i)(s_{ij}n_\gamma+1)-n_i(1-n_j)s_{ij}n_\gamma\right]\notag\\
&+\sum_{j<i}W_{ij}\left[n_j(1-n_i)s_{ij}n_\gamma-n_i(1-n_j)(s_{ij}n_\gamma+1)\right]\notag\\
&-\frac{n_i}{\tau_i},\\
\frac{dn_\gamma}{dt}&=N_{NP}\sum_{i,j>i}s_{ij}W_{ij}\left[(n_j-n_i)n_\gamma+n_j(1-n_i)\right]-\frac{n_\gamma}{\tau_\gamma},
\end{align}

where the levels are ordered in terms of increasing energy, $E_i$, and $s_{ij}=1$ if $|E_i-E_j|=E_\gamma$ and $s_{ij}=0$ otherwise. 
$N_{NP}$ is the number of nanoparticles in the THz cavity. $\tau_{\gamma}$ is the photon lifetime and $\tau_i$ the lifetime of state $i$ in the ladder. The former depends on the quality of the THz cavity while the latter will depend on the transition dipole moment of state $i$ to the surrounding states that are not a part of the ladder. In our calculation we looked at the transition rate for each state to lower states outside of the ladder and used their sum to determine the lifetime. This method gives us an overestimate of the loss since we disregard Pauli blocking (by assuming each outside state is always empty) and we disregard the process where the electron can return to a THz emitting state.

\begin{figure}
\includegraphics[width=0.5\textwidth]{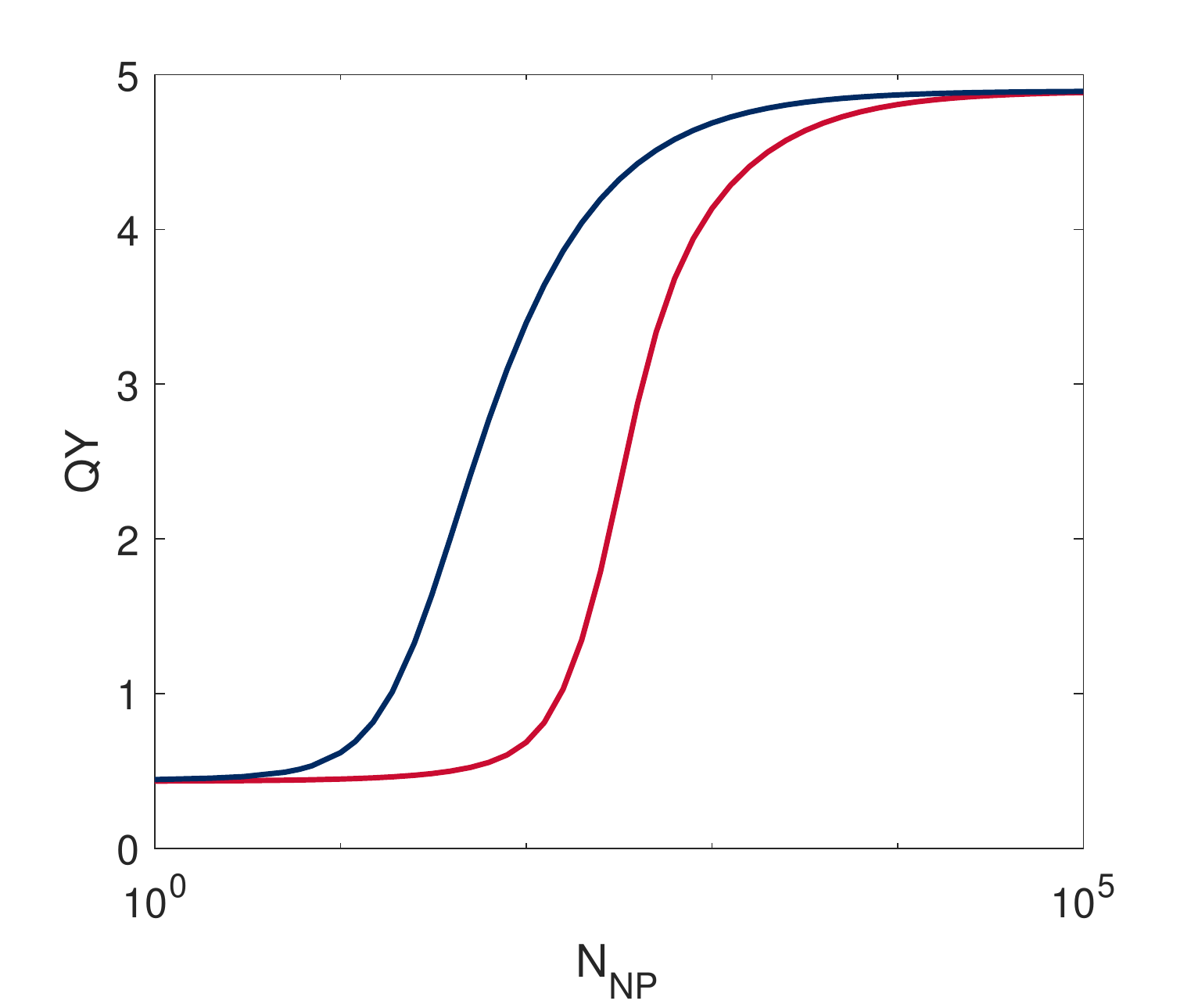}
\caption{(Color online) The relationship between the quantum yield (QY) and the number of nanoparticles in the THz cavity ($N_{NP}$). The QY describes how many THz photons we get from the system for each excitation to the highest state. The darker blue line corresponds to the QY when material profile of the nanoparticles follows the smooth $x(r)$ curve shown in figure 2 and the cavity quality factor is taken to be on the order of $10^5$. The red curve is the result when the material profile is described by the step function in the same figure. In that case the quality factor needs to be lower to accommodate for the increased linewidth. Here we take it to be $10^2$.
Both curves were calculated using the same parameters as before, nanoparticles of radius $R=63$ nm and the THz transition is $\omega_{THz}=1.1$ THz.}
\label{fig:QE}
\end{figure}

We can now use these rate equations to calculate the quantum yield (QY), or how many THz photons we produce each time we excite the nanoparticle to the highest level of the ladder, here the $6p$ state. When we consider multiple particles in the same cavity we increase the amount of THz photons and therefore increase the probability of stimulated scattering between the transitions marked with solid black arrows in Fig. \ref{fig:levels}. 
This calculation was done by adding to the coupled rate equations the mode of escaped particles, $n_{esc}(t)=\int_0^t\frac{n_\gamma(t')}{\tau_\gamma}dt'$. Now, $n_{esc}$ will tell us how many THz photons we can extract when $N_{NP}$ nanoparticles are excited, so the QY is simply $QY=\frac{n_{esc}(T)}{N_{NP}}$, where the time $T$ is taken to be $T\gg\tau_\gamma$.
Fig. \ref{fig:QE} shows how the QY depends on $N_{NP}$. The thicker blue line shows the results while using the continuous $x(r)$ we got from running our algorithm for 10 iterations (solid dark curve in the top of Fig. \ref{fig:x}) and a cavity quality factor of $10^5$. We see that the QY starts low but when the number of nanoparticles in the cavity increases it gets close to the theoretical limit of 5.
The thinner red line in Fig. \ref{fig:QE} corresponds to the results where the material profile is taken to be the step function approximating the continuous function from before. Since we consider a smaller quality factor, the QY is smaller, but still reaches almost 5 for $N_{NP}\sim 10^4$. If we assume a cavity of dimensions 1 mm $\times$ 1 mm $\times$ 1 mm that leads to a packing factor of only $10^{-8}$, which means that the distance between nanoparticles is large.


As our system essentially amounts to a gain medium, with population inversion, located in a cavity it could also be considered a THz laser. A coherent statistics would be expected of the emission \cite{Liew2017}, although we do not calculate it here, restricting ourselves to a semiclassical theory.

{\it Conclusion.---} We have proposed a new source for THz radiation: a nanoparticle with material composition changing along the radial direction. The composition is chosen in a way that gives rise to a cascade of transitions in the THz range. Using this approach we can achieve quantum yield far exceeding unity. Here we show numerical calculations for Ga$_{1-x}$Al$_x$As nanoparticles to verify our claims, where we find that the quantum yield reaches 5 for multiple nanoparticles in a terahertz cavity, which is the theoretical limit since it is the number of steps in the ladder defined by our chosen level set $\{L_i\}$. We also show that by using a step function approximation of the profile we can still achieve the same quantum yield for large number of nanoparticles in the cavity.
\\

The authors would like to thank Prof. Andrei Manolescu for useful discussions.
The work was supported by the MOE AcRF Tier 2 Grant No.
2017-T2-1-001.

\end{document}